\documentclass[twocolumn,aps,prd]{revtex4-1}
\usepackage{amsmath,graphicx,bm,subfigure}
\begin{document}
    
\title{Velocity Structure of Self-Similar Spherically Collapsed Halos}

\author{Phillip Zukin}
\email{zukin@mit.edu}
\author{Edmund Bertschinger}
\affiliation{Department of Physics, MIT, 77 Massachusetts Ave.,
Cambridge, MA 02139}

\begin{abstract}
Using a generalized self-similar secondary infall model, which accounts for tidal torques acting on the halo, we analyze the velocity profiles of halos in order to gain intuition for N-body simulation results. We analytically calculate the asymptotic behavior of the internal radial and tangential kinetic energy profiles in different radial regimes. We then numerically compute the velocity anisotropy and pseudo-phase-space density profiles and compare them to recent N-body simulations. For cosmological initial conditions, we find both numerically and analytically that the anisotropy profile asymptotes at small radii to a constant set by model parameters. It rises on intermediate scales as the velocity dispersion becomes more radially dominated and then drops off at radii larger than the virial radius where the radial velocity dispersion vanishes in our model. The pseudo-phase-space density is universal on intermediate and large scales. However, its asymptotic slope on small scales depends on the halo mass and on how mass shells are torqued after turnaround. The results largely confirm N-body simulations but show some differences that are likely due to our assumption of a one-dimensional phase space manifold. 
\end{abstract}

\pacs{04.50.Kd,04.50.-h,95.36.+x}

\maketitle

\section{Introduction}
\label{sec:intro}

Recent N-body simulations have revealed a wealth of information about the velocity structure of halos \cite{Aquarius,Ludlow,Vogelsberger}. However, simulations have finite dynamic range. Moreover, it is difficult to draw understanding from their analysis, and computational resources limit the smallest resolvable radius, since probing smaller scales require using more particles and smaller time steps. Hence, it seems natural to use analytic techniques, which do not suffer from resolution limits, to analyze the velocity distributions of halos. 

Numerous authors have analytically investigated the density profiles of halos. Work began with Gunn and Gott where they analyzed the continuous accretion of mass shells onto an initial overdensity \cite{GunnGott72,Gott75,Gunn77}. This process is known as secondary infall. By imposing that the mass accretion is self-similar, Fillmore and Goldreich \cite{FG} and Bertschinger \cite{Bert}, assuming purely radial orbits, were able to analytically calculate the asymptotic slope of the density profile. Since then, there have been numerous extensions, some which do not assume self-similarity, that take into account non-radial motions \cite{RydenGunn,Nusser,Hiotelis,WilliamsEtAl,Sikivie,DelPopolo,WhiteZaritsky,LeDelliou,Ascasibar}. Those works that do not impose self-similarity can only infer information about the velocity dispersion using the virial theorem. Hence they cannot predict a halo's velocity anisotropy. Those works that do impose self-similarity focus only on the asymptotic slopes of density profiles.  

In this paper, we analytically and numerically analyze the velocity structure of halos using an extended self-similar secondary infall model \cite{zukin}. The work which introduced this extended infall model will hereafter be referred to as Paper I. We then compare the predictions of our halo model to simulation results, focusing on the velocity anisotropy \cite{BinneyTremaine} and pseudo-phase-space density profiles \cite{TN,Aquarius,Ludlow}.  

Density profiles do not uniquely determine a self gravitating system. In order to more fully characterize dark matter halos, one needs to probe their phase space distributions. The velocity anisotropy and pseudo-phase-space density profiles are thereby useful since they complement density profiles by revealing additional information about the phase space structure of the halo.

In section \ref{sec:SSM} we summarize our generalized secondary infall model and discuss how to numerically calculate the radial and tangential velocity dispersions in the halo. In section \ref{sec:AB} we analytically calculate the asymptotic behavior of the radial and tangential kinetic energy profiles on small and intermediate scales. In section \ref{sec:Nbody} we compare our numerically calculated anisotropy and pseudo-phase-space density profiles to recent N-body results and conclude in section \ref{sec:Dis}. 

\section{Self-Similar Model}
\label{sec:SSM}

Here we first summarize the self-similar halo formation model developed in Paper I. The model is characterized by four parameters $\{n,p,B,\varpi\}$ which are described below.

In this model, the universe is initially composed of a linear spherically symmetric density perturbation with mass shells that move approximately with the hubble flow. Because of the central overdensity, mass shells eventually stop their radially outward motion and turn around. The radius at which a mass shell first turns around, or its first apocenter, is known as the turnaround radius. Since the average density is a decreasing function from the central overdensity, mass shells initially farther away will turnaround later. The halo grows by continuously accreting mass shells. Mass shells are labeled by their turnaround time $t_*$ or their turnaround radius $r_*$. 

Model parameter $n$ characterizes how quickly the initial linear density field falls off with radius ($\delta\propto r^{-n}$). It is related to the effective primordial power spectral index $n_{\rm{eff}}$ ($d\ln P/d\ln k$) through $n=n_{\rm{eff}}+3$ \cite{HoffmanShaham}. Since $n_{\rm{eff}}$ depends on scale, $n$ is set by the halo mass. As in Paper I, we restrict our attention to $0<n<3$ so that the initial density field decreases with radius while the excess mass increases with radius. Note however that $n>1.4$ corresponds to objects larger than galaxy clusters today. Model parameter $n$ also sets the growth of the current turnaround radius: $r_{ta}\propto t^{\beta}$ where $\beta=2(1+n)/3n$ \cite{zukin}.

Self-similarity imposes that at time $t$, the angular momentum per unit mass $L$ of a particle in a shell at $r$, and the density $\rho$ and mass $M$ of the halo have the following functional forms \cite{zukin}.

\begin{eqnarray}
L(r,t)&=&B\frac{r_{\rm{ta}}^2(t)}{t}f(\lambda, t/t_*)\label{LDef} \\
\rho(r,t)&=&\rho_B(t)D(\lambda)\label{rhoDef} \\
M(r,t) &=& \frac{4\pi}{3}\rho_B(t)r_{\rm{ta}}^3(t){\cal{M}}(\lambda)\label{MDef}
\end{eqnarray} 

\noindent where $\lambda\equiv r/r_{\rm{ta}}(t)$ is the radius scaled to the current turnaround radius, $\rho_B=1/6\pi Gt^2$ is the background density for an Einstein de-Sitter (flat $\Omega_m=1$) universe, and $B$ is a constant. Inspired by tidal-torque theory and numerical simulations, we take $f$ to be:

\begin{equation} \label{torque}
f(\lambda,t/t_*)= 
\begin{cases} 
\lambda^{(4-p)/2} & \text{if $t<t_*$}, \\ 
(t/t_*)^{\varpi+1-2\beta} & \text{if $t> t_*$}. 
\end{cases} 
\end{equation} 

Model parameter $p$, defined above, sets how quickly angular momentum builds up before turnaround while $B$ sets the amplitude of angular momentum at turnaround. In Paper I, using cosmological linear perturbation theory, we constrained $p$ and $B$ so that the angular momentum of particles before turnaround evolves as tidal torque theory predicts \cite{Hoyle,Peebles,White,Dorosh}. Conveniently both $p$ and $B$ are set by the halo mass. However, after comparing to density profiles from N-body simulations, we found that our expression for $B$ derived from linear theory overestimates the actual value. Hence, for the rest of this paper, the notation $B_{1.5}$ ($B_{2.3}$) signifies using a value of $B$ divided by 1.5 (2.3).

Model parameter $\varpi$, defined above, sets how quickly the angular momentum of particles grows after turnaround. This parameter is difficult to constrain analytically since the halo is nonlinear after turnaround. However, in Paper I we showed that a nonzero $\varpi$ can be sourced by substructure. Moreover, $\varpi$ influences the density profile at small scales since it controls how the pericenters of shells evolve over time. 

The trajectory of a shell after turnaround contains all of the velocity information in the halo. The trajectory's evolution equation, which follows from Newton's law, is given by:  

\begin{equation}\label{EOMSS2}
\frac{d^2\lambda}{d\xi^2}+(2\beta-1)\frac{d\lambda}{d\xi}+\beta(\beta-1)\lambda=-\frac{2}{9}\frac{{\cal{M}}(\lambda)}{\lambda^2}+\frac{B^2}{\lambda^3}e^{2(\varpi+1-2\beta)\xi}
\end{equation}

\noindent where $\xi\equiv \ln(t/t_{ta})$ and $t_{ta}$ is the current turnaround time. The initial conditions for eq.\ (\ref{EOMSS2}) are $\lambda(\xi=0)=1$ and $d\lambda/d\xi(\xi=0)=-\beta$. Calculating ${\cal{M}}(1)$ requires evolving both the shell's trajectory and ${\cal{M}}(\lambda)$ before turnaround \cite{zukin}. Because of self-similarity, the trajectory $\lambda(\xi)$ can either be interpreted as labeling the location of a particular mass shell at different times, or labeling the location of all mass shells at a particular time. We take advantage of the second interpretation in order to numerically calculate the velocity profiles. 

Inside the shell that is currently at its second apocenter, multiple shells exist at all radii. This can be seen from Figure 5 of \cite{Bert}, which plots the location of all shells at a particular time. Therefore the expectation value of a quantity $h$, for example the radial velocity, at radius $r$ and time $t$ is the value of $h$ for each shell at $r$ weighted by each shell's mass. We find:

\begin{equation}\label{ave}
\left\langle h(r,t)\right\rangle=\frac{\int^{M_{ta}}_{0}\frac{dM_*}{M_{ta}}h(t,t_*)\delta^D\big(\lambda-\lambda(\xi)\big)}{\int^{M_{ta}}_{0}\frac{dM_*}{M_{ta}}\delta^D\big(\lambda-\lambda(\xi)\big)}
\end{equation}

\noindent where $M_{ta}$ is the current turnaround mass, $h(t,t_*)$ represents the value of $h$ for the shell with turnaround time $t_*$, $dM_*$ is the mass of the shell with turnaround time $t_*$, and $\delta^D$ is the dirac delta function which picks out all shells at $r$. 

In order to numerically calculate $h$, we must relate $h(t,t_*)$ to $\lambda(\xi)$, the computed trajectory of the shell which turns around at $t_{ta}$. Using self-similarity, we find:

\begin{eqnarray}
v_t(t,t_*)&=&B\frac{r_{ta}}{t}e^{(1+\varpi-2\beta)\xi_*}\frac{1}{\lambda(\xi_*)}\label{velt} \\
v_r(t,t_*)&=&\frac{r_{ta}}{t}e^{-\beta\xi_*}\frac{d}{d\xi}\left[e^{\beta\xi}\lambda(\xi)\right]|_{\xi=\xi_*}\label{velr}
\end{eqnarray}

\noindent where $\xi_*\equiv t/t_*$ and $v_t$ ($v_r$) is the tangential (radial) velocity. Using eqs.\ (\ref{ave})-(\ref{velr}), and taking advantage of the delta function, the tangential ($\sigma^2_t$) and radial ($\sigma^2_r$) velocity dispersions become:

\begin{eqnarray}
\sigma^2_t(r,t)&\equiv&\left\langle v^2_t(r,t)\right\rangle \nonumber \\
 &=&\frac{r^2_{ta}}{t^2}\frac{\sum_ie^{(4-7\beta+2\varpi)\xi_i}\lambda^{-2}_i\left|d\lambda/d\xi\right|^{-1}_i}{\sum_ie^{(2-3\beta)\xi_i}\left|d\lambda/d\xi\right|^{-1}_i} \label{tdispersion} \\
\sigma^2_r(r,t)&\equiv&\left\langle v^2_r(r,t)\right\rangle - \left\langle v_r(r,t)\right\rangle^2 \nonumber \\
&=& \frac{r^2_{ta}}{t^2}\frac{\sum_i e^{(2-5\beta)\xi_i}\left[d(e^{\beta\xi}\lambda)/d\xi\right]^2_i\left|d\lambda/d\xi\right|^{-1}_i}{\sum_ie^{(2-3\beta)\xi_i}\left|d\lambda/d\xi\right|^{-1}_i} \nonumber \\
&-& \frac{r^2_{ta}}{t^2}\left(\frac{\sum_i e^{(2-4\beta)\xi_i}\left[d(e^{\beta\xi}\lambda)/d\xi\right]_i\left|d\lambda/d\xi\right|^{-1}_i}{\sum_ie^{(2-3\beta)\xi_i}\left|d\lambda/d\xi\right|^{-1}_i}\right)^2\nonumber \label{rdispersion}\\
\end{eqnarray} 

\noindent where $\xi_i$ is the $i$th root that satisfies $\lambda=\lambda(\xi)$. In the above, we've imposed $\left\langle v_t(r,t)\right\rangle = 0$ since our model assumes that the orbital planes of particles in a given shell are oriented in random directions. Note that inside the shell that is currently at its second apocenter, interference between multiple shells forces $\left\langle v_r(r,t)\right\rangle$ to quickly go to zero.  

\section{Asymptotic Behavior}
\label{sec:AB}

Here, using techniques developed in \cite{FG}, we analytically calculate the logarithmic slope of the tangential and radial kinetic energy in two different radial regimes. We accomplish this by taking advantage of adiabatic invariance and self consistently calculating the total radial and tangential kinetic energy profiles of the halo.   

We start by parametrizing the halo mass, radial kinetic energy $K_r$, tangential kinetic energy $K_t$ and the variation of the apocenter distance $r_a$. 

\begin{eqnarray}
M(r,t)&=&\kappa(t)r^{\alpha} \label{MassParam}\\
K_r(r,t)&=&\kappa_r(t)r^{\alpha_r} \label{KinRParam} \\
K_t(r,t)&=&\kappa_t(t)r^{\alpha_t} \label{KinTParam} \\
\frac{r_a}{r_*}&=&\left(\frac{t}{t_*}\right)^q \label{ra}
\end{eqnarray}

\noindent In the above $r_*$ is the turnaround radius of a mass shell that turns around at $t_*$. As was shown in Paper I, adiabatic invariance allows us to relate $q$ and $\alpha$ to $n$. At late times, the orbital period is much smaller than the time scale for the mass and angular momentum to grow. Integrating Newton's equation and assuming $\kappa(t)$ and $L(t)$ change little over an orbit, we find:

\begin{equation}\label{energy}
\left(\frac{dr}{dt}\right)^2= \frac{2G\kappa(t)}{\alpha-1}(r_a^{\alpha-1}-r^{\alpha-1})-L^2(t)(r^{-2}-r_a^{-2}).   
\end{equation}

\noindent The above relationship tell us how the pericenters $r_p$ evolve with time. Defining $y\equiv r_a/r_p$ and evaluating the above at $r=r_p$, we find:

\begin{equation}\label{peri}
\frac{1-y^{\alpha-1}}{y^{-2}-1}\equiv A(y) =\frac{(\alpha-1)L^2(t)}{2G\kappa(t)r_a^{\alpha+1}(t)}
\end{equation}

In Paper I, by analyzing the radial action, we found that when $y\ll1$, $\kappa(t)r^{\alpha+1}_a(t)=\rm{const}$. Therefore,  for $\varpi<0$, eq.\ (\ref{peri}) implies that $y$ will decrease over time. However, for $\varpi>0$, $\kappa(t)r^{\alpha+1}_a(t)=\rm{const}$ and eq.\ (\ref{peri}) imply that $y$ will increase over time and will at one point violate $y\ll1$. Since we only consider bound orbits, the constraint $y\leq1$ holds. At late times, as the angular momentum continues to increase for $\varpi>0$, $y\sim1$, orbits become approximately circular, the radial action vanishes, and $L^2(t)\sim\kappa(t)r^{\alpha+1}_a(t)$. Hence halos with $\varpi<0$ will have orbits that become more radial over time ($y\ll1$) while halos with $\varpi>0$ will have orbits that become more circular over time ($y\sim1$). The above insight leads to the following constraint.

\begin{equation}\label{q}
q=
\begin{cases}
\frac{1}{\alpha+1}\{2\varpi+\frac{2}{3n}[\alpha(1+n)-3]\} & \text{if $\varpi\geq0$} \\
\frac{2}{3n(\alpha+1)}\left[\alpha(1+n)-3\right]  & \text{if $\varpi<0$}
\end{cases}
\end{equation} 

\noindent For the specific case, $\varpi<0$, taking advantage of $y\ll1$, the adiabatic invariance arguments above, and eqs.\ (\ref{LDef}) and (\ref{torque}), we can rewrite eqn.\ (\ref{peri}) in the form $y(t,t_*)=y_0(t/t_*)^l$, where: 

\begin{equation} \label{peri2}
l= 
\begin{cases} 
\varpi & \text{if $\alpha>1$}, \\ 
2\varpi/(\alpha+1) & \text{if $\alpha<1$}. 
\end{cases} 
\end{equation} 

\noindent and $y_0r_{*}$ is the pericenter of a mass shell at turnaround. The special case $\alpha=1$ will be addressed later.

We next calculate the kinetic energy profiles. After a few orbits, shells oscillate at a much higher frequency than the growth rate of the halo. When calculating the internal mass profile, this allows us to weight each mass shell based on how much time it spends interior to a certain scale \cite{FG}. Likewise, when calculating the total internal kinetic energy profile, we can weight each mass shell by both a time-averaged $v^2_t$ (or $v^2_r$) and a factor that accounts for how often the shell lies interior to a certain scale. For a derivation, please see the Appendix. 

Using eqs.\ (\ref{energy}) and (\ref{peri}), the kinetic energy weighting $P_i(r/r_a,y)$ at time $t$ for a mass shell with apocenter distance $r_a$, pericenter $yr_a$, below $r$, is:

\begin{eqnarray}
P_i(u,y)&=&0\quad\quad\quad\;\;\;\;\;\; (u < y) \nonumber \\
P_i(u,y)&=&\frac{I_i(u,y)}{I(1,y)} \quad (y< u\leq1) \nonumber \\
P_i(u,y)&=&\frac{I_i(1,y)}{I(1,y)}\quad\;\;\; (u>1)
\end{eqnarray}

\noindent where 

\begin{eqnarray}
I_t(u,y)&=& \frac{1}{2}\left(\frac{Br^2_*}{t_*r_a}\right)^2\left(\frac{t}{t_*}\right)^{2\varpi}\int^u_y\frac{dv}{v^2f(v,y)} \\
I_r(u,y)&\propto& \frac{r^2_{ta}}{t^2}\left(\frac{r_a}{r_{ta}}\right)^{\alpha-1}\int^u_yf(v,y)dv \label{IR} \\
I(u,y)&=& \int^u_y\frac{dv}{f(v,y)}
\end{eqnarray}

\noindent and

\begin{equation} \label{fvy}
f(v,y)\equiv
\begin{cases}
\left((1-v^{\alpha-1})-A(y)(v^{-2}-1)\right)^{1/2} & \text{if $\alpha>1$}, \\ 
\left((v^{\alpha-1}-1)+A(y)(v^{-2}-1)\right)^{1/2} & \text{if $\alpha<1$}.
\end{cases}
\end{equation}

\noindent The index $i=\{r,t\}$ is used for shorthand to represent either the radial or tangential direction and the dependence of $r_a$ and $r_p$ on $t_*$ is implicit. Eq.\ (\ref{IR}) is not an equality since $G\kappa(t) \propto r^{3-\alpha}_{ta}/t^2$. Moreover, the proportionality constant varies for different radial regimes in the halo. 

Similar to the treatment in Paper I, self consistency demands that:

\begin{eqnarray}\label{consistency}
\left(\frac{r}{r_{ta}}\right)^{\alpha_i}&=&\frac{K_i(r,t)}{K_i(r_{ta},t)}\nonumber \\
&\propto&\int^{M_{ta}}_0\frac{dM_*}{M_{ta}}\frac{t^2}{r^2_{ta}}P_i\left(\frac{r}{r_a(t,t_*)},y(t,t_*)\right)\nonumber \\
\end{eqnarray} 

\noindent where $dM_*$ is the mass of a shell that turned around at $t_*$ and $M_{ta}$ is the current turnaround mass. The above is not an equality, even for the tangential kinetic energy, because of a proportionality constant, similar to ${\cal{M}}(1)$, that is not included. See the Appendix for details. Its numerically computed value does not affect the asymptotic slopes $\alpha_i$. Noting from eq.\ (\ref{MDef}) that

\begin{equation}
\frac{dM_*}{d \ln t_*}=(3\beta-2)M_{ta}\left(\frac{t}{t_*}\right)^{3\beta-2},
\end{equation}

\noindent using eq.\ (\ref{ra}) and transforming integration variables to $u\equiv r/r_a$, we find:

\begin{equation}\label{Consistency2}
\left(\frac{r}{r_{ta}}\right)^{\alpha_i-k}\propto \frac{t^2}{r^2_{ta}}\int^{\infty}_{r/r_{ta}}\frac{du}{u^{1+k}}P_i(u,y(t,t_*))
\end{equation}

\noindent where $k=(3\beta-2)/(\beta-q)$. As $u$ increases, the above integral sums over shells with smaller $t_*$. Since the pericenter of a shell evolves with time, the second argument of $P_i$ depends on $u$. The dependence varies with torque model (sign of $\varpi$); hence we've kept the dependence on $u$ implicit. Next we analyze the above for certain regimes of $r/r_{ta}$, and certain torquing models, in order to constrain the relationship between $\alpha_i$ and $k$. 

\subsection{$r/r_{\rm{ta}} \ll y_0$, $\varpi<0$}

For $\varpi<0$, particles lose angular momentum over time. When probing scales $r/r_{\rm{ta}}\ll y_0$, mass shells with $t_*\ll t_{ta}$ only contribute. As a result, $y(t,t_*)\ll1$. Using eq.\ (\ref{peri2}), we then find:

\begin{equation}\label{peri3}
y(t,t_*)=y_0\left(\frac{t}{t_*}\right)^l=y_0\left(\frac{r}{ur_{\rm{ta}}}\right)^\delta
\end{equation}

\noindent where $\delta\equiv l/(q-\beta)$ and $u\equiv r/r_a$. For bound mass shells, $q-\beta<0$. Therefore, since $\delta>0$, the first argument of $P_i$ in eq.\ (\ref{Consistency2}) increases while the second decreases as we sum over shells that have turned around at earlier and earlier times ($u\rightarrow\infty$). For $r/r_{\rm{ta}} \ll y_0$, mass shells which most recently turned around do not contribute to the kinetic energy inside $r/r_{\rm{ta}}$ since we are probing scales below their pericenters. Mass shells only begin to contribute when the two argument of $P_i$ are roughly equal to each other. This occurs around:

\begin{equation}\label{yprime}
u = y_1\equiv \left(y_0(r/r_{\rm{ta}})^{\delta}\right)^{1/(1+\delta)}
\end{equation}

\noindent Hence, we can replace the lower limit of integration in eq.\ (\ref{Consistency2}) with $y_1$. We next want to calculate the behavior of eq.\ (\ref{Consistency2}) close to $y_1$ in order to determine whether the integrand is dominated by mass shells around $y_1$ or mass shells that have turned around at much earlier times. The first step is to calculate the behavior of $P_i(u,y)$ for $u\approx y$. We find:

\begin{eqnarray}
P_t(u,y)&\propto& \frac{r^2_{ta}}{t^2}\left(\frac{r}{ur_{ta}}\right)^{l_t}u^{1/2}(1-y/u)^{1/2} \nonumber\\
&\times& 
\begin{cases} y^{-3/2} & \text{if $\alpha>1$}, \\ 
y^{-1-\alpha/2} & \text{if $\alpha<1$}.
\end{cases} \\
P_r(u,y)&\propto& \frac{r^2_{ta}}{t^2}\left(\frac{r}{ur_{ta}}\right)^{l_r}u^{3/2}(1-y/u)^{3/2} \nonumber\\
&\times& 
\begin{cases} y^{-1/2} & \text{if $\alpha>1$}, \\ 
y^{-1+\alpha/2} & \text{if $\alpha<1$}.
\end{cases}
\end{eqnarray}

\noindent where $l_t=2(1+\varpi-q-\beta)/(q-\beta)$ and $l_r=\alpha-1$. Given the above, we evaluate the indefinite integral in eq.\ (\ref{Consistency2}), noting that $y$ is a function of $u$ (eq.\ {\ref{peri3}}). For $u\sim y_1$, we find:

\begin{eqnarray}
&&\frac{t^2}{r^2_{ta}}\int\frac{du}{u^{1+k}}P_t\left(u,y_0\left(\frac{r}{ur_{\rm{ta}}}\right)^{\delta}\right) \nonumber \\
&\propto& (u/y_1-1)^{3/2}\left(\frac{r}{r_{ta}}\right)^{l_t}\times
\begin{cases}
y_1^{-1-k-l_t} & \text{if $\alpha>1$}, \\
y_1^{-1/2-k-l_t-\alpha/2} & \text{if $\alpha<1$}. \nonumber 
\end{cases} \\ \label{integral1} \\
&&\frac{t^2}{r^2_{ta}}\int\frac{du}{u^{1+k}}P_r\left(u,y_0\left(\frac{r}{ur_{\rm{ta}}}\right)^{\delta}\right) \nonumber \\
&\propto& (u/y_1-1)^{5/2}\left(\frac{r}{r_{ta}}\right)^{l_r}\times
\begin{cases}
y_1^{1-k-l_r} & \text{if $\alpha>1$}, \\
y_1^{1/2-k-l_r+\alpha/2} & \text{if $\alpha<1$}. \nonumber 
\end{cases} \label{integral2} \\
\end{eqnarray} 

Following the logic in Paper I, if we keep $u/y_1$ fixed and the integrand blows up as $y_1 \rightarrow 0$, then the left hand side of eq.\ (\ref{Consistency2}) must diverge in the same way as the right hand side shown in eqs.\ (\ref{integral1}) and (\ref{integral2}). Therefore, using eq.\ (\ref{yprime}):

\begin{eqnarray}
\alpha_t-k-l_t&=& 
\begin{cases}
-\delta(1+k+l_t)/(1+\delta)& \text{if $\alpha>1$}, \\ 
-\delta(1/2+k+l_t+\alpha/2)(1+\delta) & \text{if $\alpha<1$}.
\end{cases} \nonumber \\ \label{diverge1} \\
\alpha_r-k-l_r&=& 
\begin{cases}
\delta(1-k-l_r)/(1+\delta)& \text{if $\alpha>1$}, \\ 
\delta(1/2-k-l_r+\alpha/2)(1+\delta) & \text{if $\alpha<1$}.
\end{cases} \nonumber \label{diverge2} \\
\end{eqnarray} 

\noindent Otherwise, if the integrand converges, then the right hand side is proportional to $(r/r_{ta})^{l_i}$. Therefore, the left hand side must also have the same scaling, which implies $\alpha_i=k+l_i$. Solving the above system of equations for $\alpha_i$ simplifies dramatically since we have already solved for $\{\alpha,k,q\}$ in Paper I. Rewritten below for convenience, we found: 

\begin{eqnarray}\label{relations1}
\text{For }n&\leq&2: \nonumber \\
\alpha&=&\frac{1+n-\sqrt{(1+n)^2+9n\varpi(n\varpi-2)}}{3n\varpi} \nonumber \\
k&=&\frac{1+n+3n\varpi-\sqrt{(1+n)^2+9n\varpi(n\varpi-2)}}{n\varpi(4+n)} \nonumber \\
q&=&\frac{1+n-3n\varpi-\sqrt{(1+n)^2+9n\varpi(n\varpi-2)}}{3n} \nonumber \\ 
\text{For }n&\geq&2: \nonumber \\
\alpha&=&k=\frac{3}{1+n}\;, \quad q=0.
\end{eqnarray}

\noindent Using eq.\ (\ref{relations1}) to solve for $\alpha_i$ and making sure the solution is consistent, (ie: using eqs.\ (\ref{diverge1}) and (\ref{diverge2}) only if the integrand diverges as $y_1\rightarrow0$), we find:

\begin{eqnarray}\label{Vrelations1}
\text{For }n&\leq&2: \nonumber \\
\alpha_t&=& \alpha_r = \frac{12-9n\varpi-2\sqrt{(1+n)^2+9n\varpi(n\varpi-2)}}{1+n+\sqrt{(1+n)^2+9n\varpi(n\varpi-2)}} \nonumber \\
\text{For }n&\geq&2: \nonumber \\
\alpha_t&=&\frac{(4+n)(3n\varpi+2n-10)}{2(1+n)(3n\varpi-n-4)} \nonumber \\
\alpha_r&=& \frac{5-n}{1+n}.
\end{eqnarray}

\noindent The above solutions are continuous at $n=2$. Taking the no-torque limit ($\varpi\rightarrow0$), we find $\alpha_t=\alpha_r=(5-n)/(1+n)$ for all $n$. Assuming virial equilibrium, one would predict $\alpha_i = 2\alpha-1=(5-n)/(1+n)$. Hence, in the no-torque limit, both the radial and tangential kinetic energy are virialized. However, when $\varpi<0$, only the radial kinetic energy for $n\geq2$ is virialized. All other profiles are out of virial equilibrium because they are dominated by shells which recently turned around and hence have not had time to virialize. Since all collapsed objects today have $n<2$, this model predicts unvirialized halos when particles lose angular momentum after turnaround.

\subsection{$r/r_{\rm{ta}} \ll y_0$, $\varpi>0$}

For $\varpi>0$, the angular momentum of particles increase with time. As mentioned above, when probing scales $r/r_{\rm{ta}} \ll y_0$, mass shells with $t_* \ll t_{ta}$ only contribute. As a result, $y(t,t_*)\sim1$. In other words, the orbits are roughly circular. We can therefore replace the lower limit of integration in eq.\ (\ref{Consistency2}) with 1 since mass shells will only start contributing to the sum when $u\sim y\sim1$. Hence, the right hand side of eq.\ (\ref{Consistency2}) is proportional to $(r/r_{ta})^{l_i}$, which implies $\alpha_i=k+l_i$. Using the results from Paper 1, reproduced below for convenience,

\begin{equation}\label{relations3}
\alpha=k=\frac{3}{1+n-3n\varpi}\;, \quad q=2\varpi\;, \quad\; \text{for}\;\; 0\leq n\leq3,
\end{equation} 

\noindent we find:

\begin{equation}\label{Vrelations3}
\alpha_t=\alpha_r=\frac{5-n+3n\varpi}{1+n-3n\varpi}\;,  \quad\quad\; \text{for}\;\; 0\leq n\leq3.
\end{equation} 

The no-torque case, $\varpi=0$, is consistent with the analysis in the prior subsection. The singularity $\varpi=(1+n)/3n$, as discussed in Paper I, corresponds to orbits that are not bound. Hence we only consider $\varpi<(1+n)/3n$. Eq.\ (\ref{Vrelations3}) shows that the halo, for $\varpi\geq0$, is in virial equilibrium ($\alpha_i=2\alpha-1$). This is expected since the velocity profiles are dominated by mass shells that have turned around at $t\ll t_{ta}$. 

\subsection{$y_0\ll r/r_{\rm{ta}} \ll1$}

In this regime, we are probing scales larger than the pericenters of the most recently turned around mass shells. As a result, $P_i(u,y)$ is dominated by the contribution from the integrand when $u\gg y$. Therefore: 

\begin{eqnarray}
P_t(u,y)&\propto& \frac{r^2_{ta}}{t^2}\left(\frac{r}{ur_{ta}}\right)^{l_t} \times
\begin{cases} y^{-1} & \text{if $\alpha>1$}, \\ 
y^{-(\alpha+1)/2} & \text{if $\alpha<1$}.
\end{cases} \nonumber \\ \\
P_r(u,y)&\propto& \frac{r^2_{ta}}{t^2}\left(\frac{r}{ur_{ta}}\right)^{l_r} \times
\begin{cases} u & \text{if $\alpha>1$}, \\ 
u^{(\alpha+1)/2} & \text{if $\alpha<1$}.
\end{cases} \nonumber \\
\end{eqnarray}

\noindent Plugging in the above into eq.\ (\ref{Consistency2}), using the results of Paper I shown below for convenience,

\begin{eqnarray}\label{relations2}
\alpha&=&1\;, \quad k=\frac{6}{4+n}\;, \quad q=\frac{n-2}{3n}\;, \quad \text{for}\; n\leq2 \nonumber \\
\alpha&=&k=\frac{3}{1+n}\;, \quad q=0\;, \quad\quad\quad\quad\quad\; \text{for}\; n\geq2 \nonumber \\
\end{eqnarray}

\noindent and utilizing the same divergence and convergence arguments above, we find:

\begin{eqnarray}\label{Vrelations2}
\text{For }n&\leq&2: \nonumber \\
\alpha_t&=& 0, \quad\quad \alpha_r= 1 \nonumber \\
\text{For }n&\geq&2: \nonumber \\
\alpha_t&=&
\begin{cases} 0 & \text{if $\varpi<\frac{5-n}{3n}$}, \\ 
\frac{5-n-3n\varpi}{1+n} & \text{if $\frac{5-n}{3n}\leq\varpi<\frac{1+n}{3n}$}.
\end{cases} \nonumber \\
\alpha_r&=& \frac{5-n}{1+n}.
\end{eqnarray}

The above solutions are continuous at $n=2$. The upper limit $(1+n)/3n$ on $\varpi$ enforces that orbits are bound ($\varpi<\beta/2$). For $n\leq2$, $\varpi>(5-n)/3n$ results in unbound orbits and hence is not considered. The radial kinetic energy follows the same profile expected from virial equilibrium ($\alpha_r=2\alpha-1$), even though recently turned around mass shells dominate the kinetic energy for $n<2$. We believe this is a result of angular momentum not playing a dynamical role at these scales. Using this logic, and as eq.\ (\ref{Vrelations2}) reveals, this is consistent with the the tangential kinetic energy not being virialized ($\alpha_t\neq2\alpha-1$). Taking the limit of eq.\ (\ref{Vrelations1}) as $\varpi\rightarrow-\infty$, we recover the same expressions as eq.\ (\ref{Vrelations2}) for $\alpha_r$. This is expected since in this limit, particles lose their angular momentum instantly, resulting in purely radial orbits. We do not expect to recover the same expressions for $\alpha_t$ since the tangential kinetic energy vanishes in this limit. 

Eqs.\ (\ref{relations1}), (\ref{relations3}) and (\ref{relations2}) determine how quickly the mass inside a fixed radius grows as a function of time ($\kappa(t)$ defined in eq.\ \ref{MassParam}). When $q=0$, apocenters of mass shells settle down to a constant fraction of their turnaround radii, leading to a constant mass inside a fixed radius. For cosmologically relevant structures ($n<1.4$), this occurs on small scales for $\varpi=0$. When $q<0$, inward migration leads to an increasing mass inside a fixed radius. When $q>0$, outward migration leads to a decreasing mass inside a fixed radius.

This section assumed $\alpha\neq1$ and yet, for certain parts of parameter space,  eqs.\ (\ref{relations1}), (\ref{relations3}) and (\ref{relations2}) give $\alpha=1$. However, since the solutions are continuous as $\alpha\rightarrow1$ from the left and right, then the results hold for $\alpha=1$ as well.

\section{Comparison with N-body Simulations}
\label{sec:Nbody}

In this section, using the analytic results derived above to gain intuition, we first analyze how $\varpi$ influences the anisotropy and pseudo-phase-space density profiles and then compare our numerically computed profiles to recent N-body simulations of galactic size halos \cite{Aquarius}.

As described in Paper I, the mass of a halo is not well defined when our model is applied to cosmological structure formation since it is unclear how the spherical top hat mass which characterizes the halo when it is linear relates to the virial mass which characterizes the halo when it is nonlinear. For halos today with galactic size virial masses, we assume the model parameter $n$ which characterizes the initial density field, is set by a spherical top hat mass of $10^{12}M_{\odot}$. Specifying the top hat mass also sets model parameters $B$ and $p$. For explicit expressions used to calculate model parameters $n$, $p$, and $B$, please see Paper I.

When analyzing the influence of $\varpi$, we use model parameters: $n=0.77$, $p=2n$, and $B_{1.5}=0.39$. When comparing to N-body simulations, we use model parameters $n=0.77$, $p=2n$, $\varpi= 0.12$, and $B_{1.5}\; (B_{2.3}) = 0.39\; (0.26)$. This value of $\varpi$ ensures $\rho\propto r^{-1}$ on small scales and, as shown in Paper I, this range in $B$ gives good agreement with the Einasto and NFW profiles \cite{NFW}. 

For the N-body comparisons, we average $\rho$, $\sigma^2_t$, and $\sigma^2_r$ in 50 spherical shells equally spaced in $\log_{10}r$ over the range $1.5\times 10^{-4} < r/r_v < 3$, where $r_v$ satisfies $M(r_v,t)=800\pi r_v^3\rho_B(t)/3$. This is the same procedure followed with the recent Aquarius simulation \cite{Aquarius}. We also calculate $r_{-2}$, the radius where $r^2\rho$ reaches a maximum. This radius, as well as the virial radius $r_v$, are commonly referred to in simulation papers. As discussed in Paper I, the density profile is isothermal for our halo over a range of $r$. Moreover, the maximum peaks associated with the caustics are unphysical. So, we choose a value of $r_{-2}$ in the isothermal regime that gives good agreement with empirical density profiles. Changing $r_{-2}$ does not change our interpretation of the results. We find $r_{-2}/r_{ta}$ = 0.07 (0.05) for $B_{1.5}$ ($B_{2.3}$). For reference, we find the dimensionless radius of first pericenter passage ($y_0$) to be 0.042 (0.026) for $B_{1.5}$ ($B_{2.3}$).

As mentioned previously, N-body simulations have finite dynamic range. The innermost radius where the simulation results can be trusted is set by the total number of particles \cite{Power}. The recent Aquarius simulations characterize their innermost radius based on the convergence of the circular velocity, at a particular radius, for the same halo simulated at different resolutions \cite{Aquarius}. The notation $r^{(1)}_{\rm{conv}}$  $(r^{(7)}_{\rm{conv}})$ corresponds to the smallest radius such that the circular velocity has converged to $10\%$ $(2.5\%)$ or better at larger radii. When these radii are showed in the figures, we use the values quoted in Table 2 of \cite{Aquarius} for halo Aq-A-2 ($r^{(1)}_{\rm{conv}}/r_{-2}=0.022$ and $r^{(7)}_{\rm{conv}}/r_{-2}=0.052$) since all six halos were simulated at this resolution.  

\subsection{Anisotropy Profile}
\label{sec:Ani}

Here we analyze the velocity anisotropy $\beta_v\equiv 1-\sigma^2_t/2\sigma^2_r$ for galactic size halos, where the tangential and radial velocity dispersions are defined in eqs.\ (\ref{tdispersion}) and (\ref{rdispersion}) respectively. Based on the analysis in Section \ref{sec:AB}, we expect $\beta_v$ to asymptote to a constant for $r/r_{ta}\ll y_0$ since $\sigma_t^2\propto \sigma^2_r$ and $\beta_v$ to increase for $y_0\ll r/r_{ta}\ll r_v/r_{ta}$ since $\sigma_t^2/\sigma_r^2 \propto r^{-1}$. Moreover, for radii larger than the first shell crossing ($r\sim r_v$), $\sigma^2_r=0$ since only one shell contributes to the dispersion. Hence, in this radial range, $\beta_v=-\infty$.  

\begin{figure}
  \begin{center}
    \includegraphics[height = 70mm, width=\columnwidth]{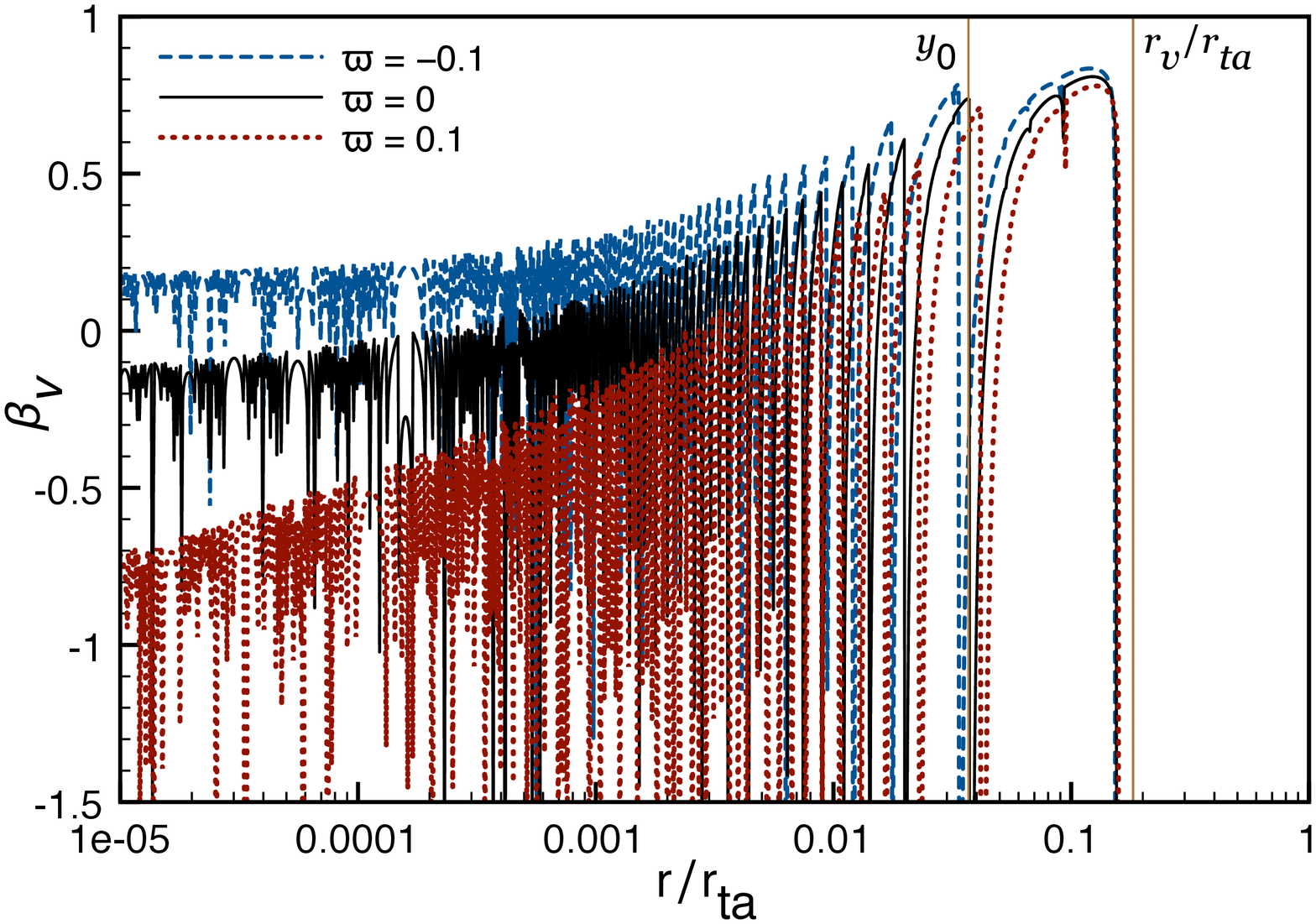}
     \includegraphics[height = 70mm, width=\columnwidth]{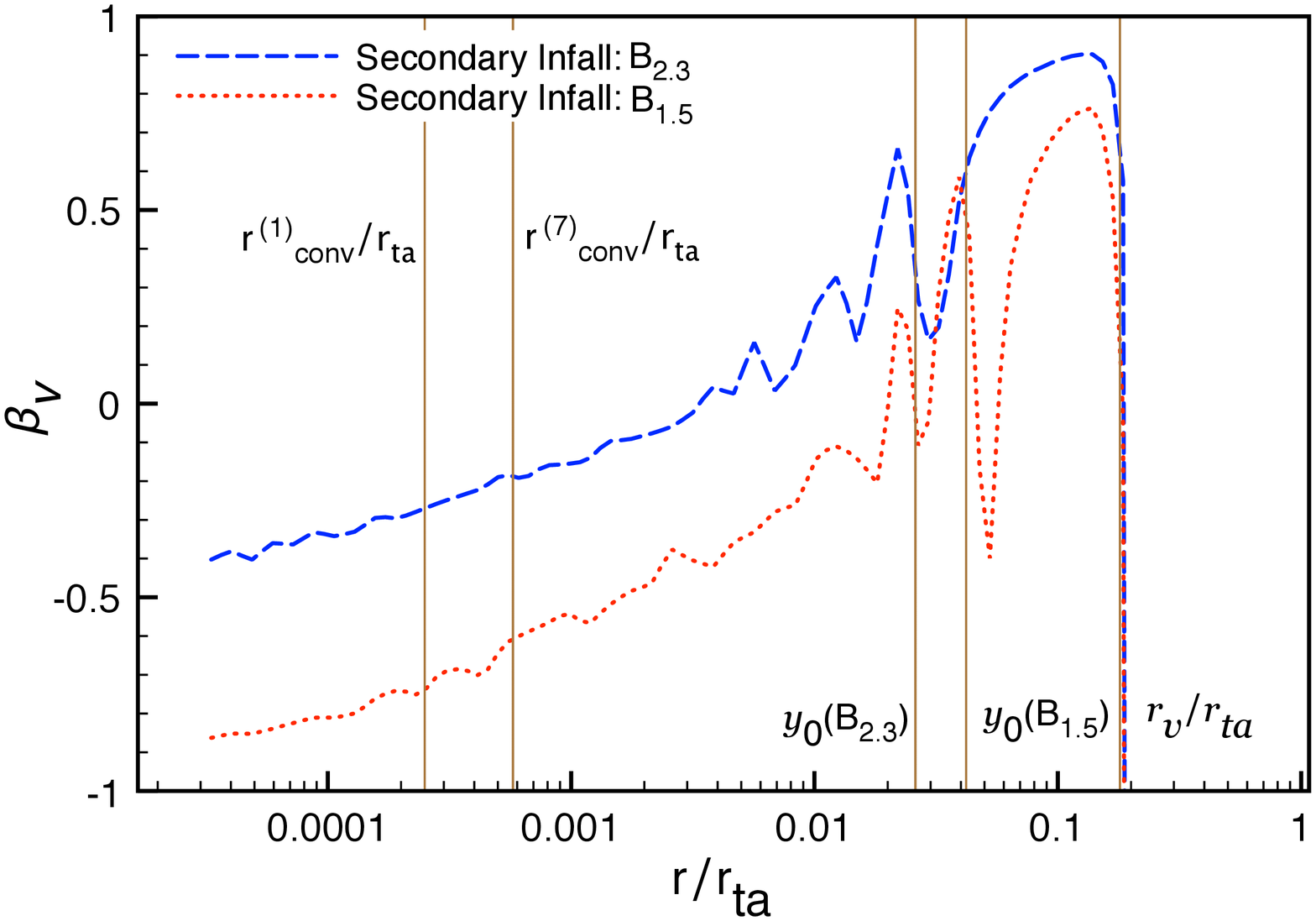}
  \end{center}
\caption{The top panel shows the velocity anisotropy profile for a self-similar halo with model parameters  $n=0.77$, $p=2n$, $B_{1.5}=0.39$, and varying $\varpi$. Smaller $\varpi$ leads to halos with more radial orbits at a particular radius. The bottom panel shows the smoothed velocity anisotropy profile for a self-similar halo with model parameters $n=0.77$, $p=2n$, $\varpi= 0.12$, and $B_{1.5}\; (B_{2.3}) = 0.39\; (0.26)$. Smaller $B$ leads to a larger peak width and more radial orbits. The profile is qualitatively similar to results from N-body simulations. The dimensionless radius of first pericenter passage $(y_0)$ and the virial radius $(r_v)$ are labeled for clarity. The convergence radii for the Aquarius halo Aq-A-2 \cite{Aquarius} are labeled for reference.\label{fig:Ani}}
\end{figure}

In the top panel of Figure \ref{fig:Ani} we plot the velocity anisotropy for galactic size halos with varying $\varpi$. In the bottom panel, we plot the smoothed velocity anisotropy for model parameters that give good agreement with density profiles from simulated galactic size halos. The downward spikes in both panels are caustics which exist because of unphysical radially cold initial conditions. In both panels, as analytically predicted, the velocity anisotropy asymptotes at small radii, increases at intermediate radii, and then drops off near the virial radius. 

The top panel shows that $\varpi$ affects the radius of first pericenter passage ($y_0$), the amplitude of $\beta_v$ close to the virial radius, as well as the asymptotic value of $\beta_v$ at small radii. This behavior is intuitive since smaller values of $\varpi$ give rise to halos populated with less circularized orbits at a given radius. Note however that the envelope of the anisotropy profile begins to increase and become more radially dominated for $r/r_{ta}<y_0$, contradicting our analytic analysis. More specifically, for $\varpi=-0.1$, $\beta_v\sim0.2$ for $r/r_{ta}<0.001$, and starts to increase at $r/r_{ta}\sim0.001$ when it should start increasing at $r/r_{ta}\sim y_0$, according to Section \ref{sec:AB}. This is a result of assumptions used to calculate $\alpha_r$ breaking down. This is more apparent for $\varpi>0$ since the orbital period is longer. However, as $r\rightarrow0$, the assumptions become more valid. 

In the bottom panel, we show how model parameter $B$ affects the velocity anisotropy. As discussed in Paper I, smaller $B$ leads to orbits that take longer to circularize and density profiles with a larger isothermal region (smaller $y_0$). The bottom panel should be compared to Figures 9 and 10 of \cite{Aquarius}. Though our model cannot address structure outside $r_v$, the graphs are qualitatively very similar. The width of the peak predicted in our model agrees with results from N-body simulations. This should be expected since the parameter $B$ was chosen so that the width of the isothermal region in the density profiles agree. However, our model over predicts the velocity anisotropy close to $r_v$ and under predicts the velocity anisotropy at small radii. In other words, at large scales the halo is populated with too many radial orbits while on small scales the halo is populated with too many circular orbits. 

\begin{figure}
  \begin{center}
    \includegraphics[height = 70mm, width=\columnwidth]{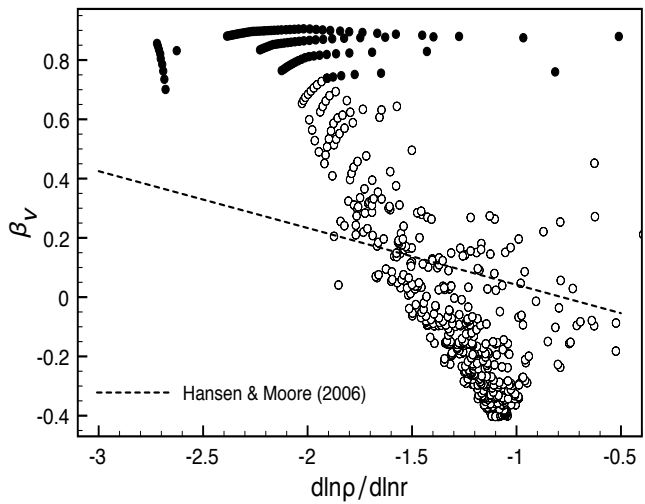}
  \end{center}
\caption{The local logarithmic slope of the density profile plotted against the velocity anisotropy. The relationship relating these two quantities that was proposed by Hansen $\&$ Moore (2006) is also showed. Open circles correspond to $1.5\times10^{-4}\;r_v<r<r_{-2}$ while closed circles correspond to $r_{-2}<r<r_{200}$. Unlike N-body simulations, our self-similar model does not fit the trend proposed by Hansen $\&$ Moore for $r<r_{-2}$. This reveals a shortcoming of the model.\label{fig:HM}}
\end{figure}

This trend is most clearly seen in Figure \ref{fig:HM}. There we plot the local velocity anisotropy versus the logarithmic slope of the density profile for a galactic size halo with $B_{2.3}=0.26$ as well as a universal relationship relating these two quantities that was derived by Hansen $\&$ Moore \cite{HM}. The open circles correspond to $1.5\times10^{-4}\;r_v<r<r_{-2}$ while the filled circles correspond to $r_{-2}<r<r_v$. This figure should be compared to Figure 11 of \cite{Aquarius}. In the Aquarius simulation paper, the Hansen $\&$ Moore prediction agrees well with N-body results for $r<r_{-2}$. However, in our Figure \ref{fig:HM}, while there is a clear trend between the local velocity anisotropy and the logarithmic slope of the density profile, that trend does not match the derived relationship. Note though that our model, just as the Aquarius simulation claimed, does show deviations from the Hansen $\&$ Moore trend for $r_{-2}<r<r_{v}$. In our model, this deviation is caused by a vanishing radial velocity dispersion. For simulated halos, other effects like non-sphericity or non-self-similarity may also play a role.   

The self-similar model's inability to match the amplitude of the velocity anisotropy seen in N-body simulations reveals a weakness in the model. Clearly, it is unphysical for all particles in a particular shell to have the same amplitude of angular momentum and the same radial velocity. In reality, a given shell should have a radial velocity dispersion and should have a distribution of angular momentum that evolves with time. This possibility will be discussed again in Section \ref{sec:Dis}.  

\subsection{Pseudo-Phase-Space Density Profile}
\label{sec:Phase}

\begin{figure*}
  \begin{center}
    \includegraphics[height = 70mm, width=\columnwidth]{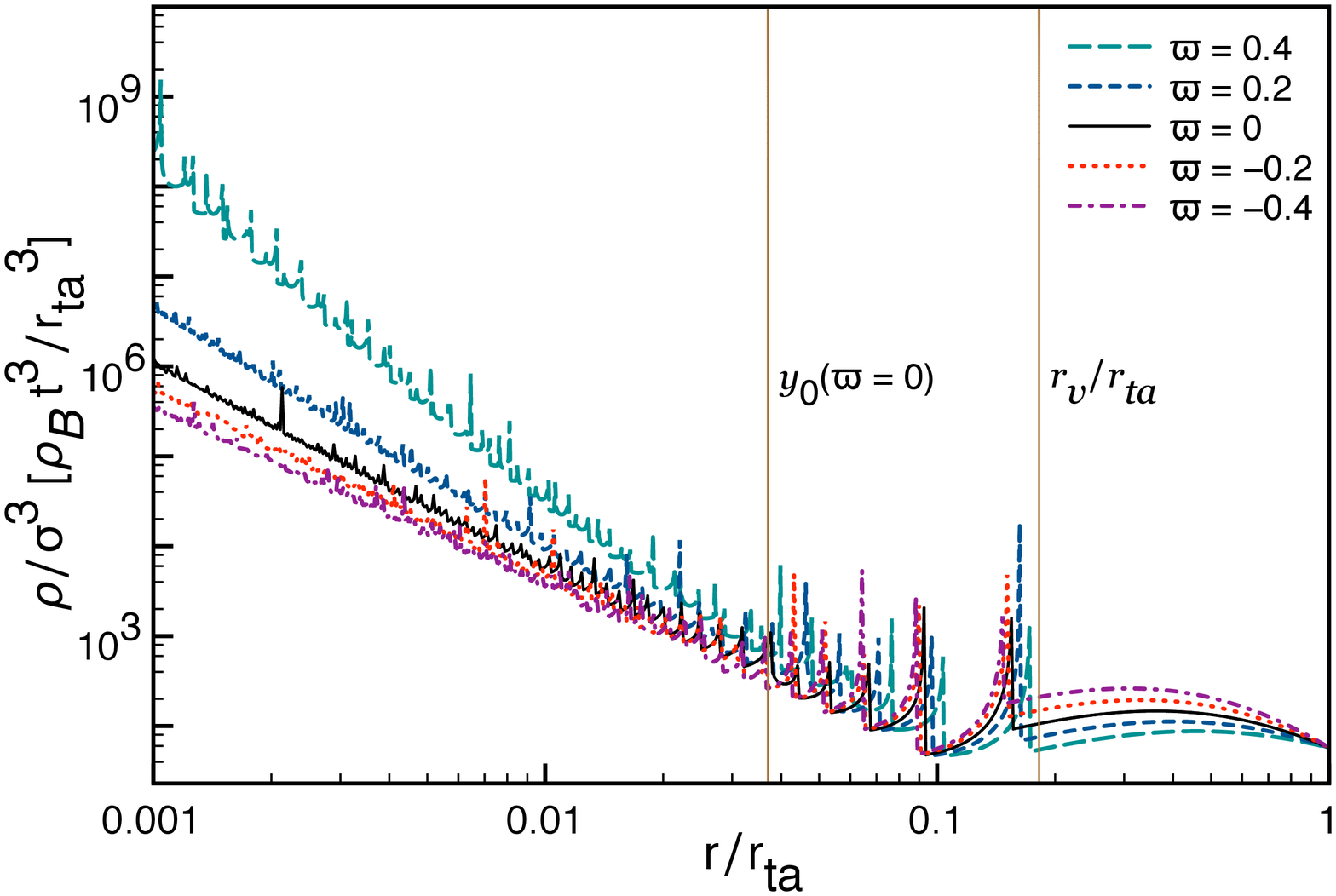}
     \includegraphics[height = 70mm, width=\columnwidth]{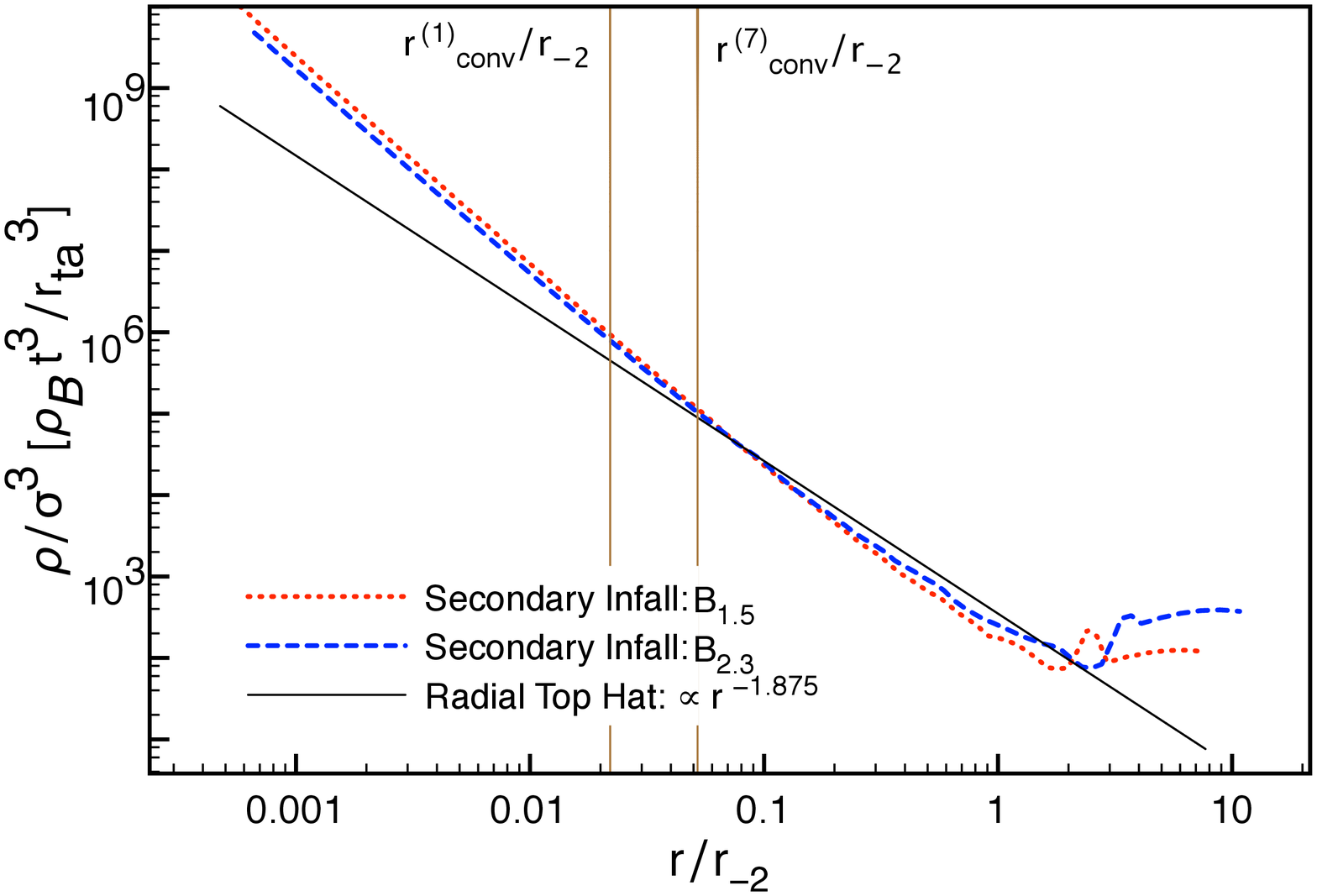}
      \includegraphics[height = 70mm, width=\columnwidth]{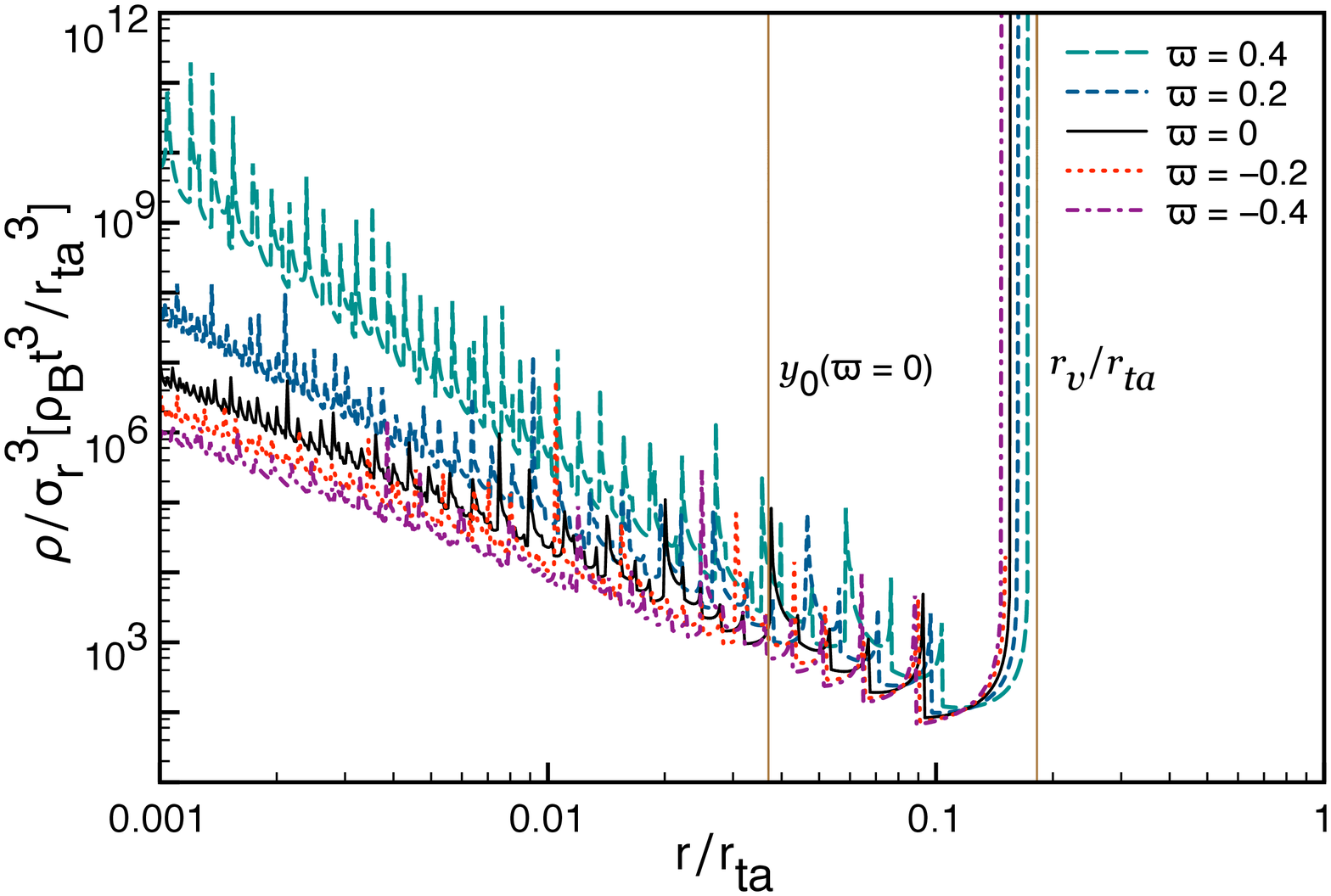}
     \includegraphics[height = 70mm, width=\columnwidth]{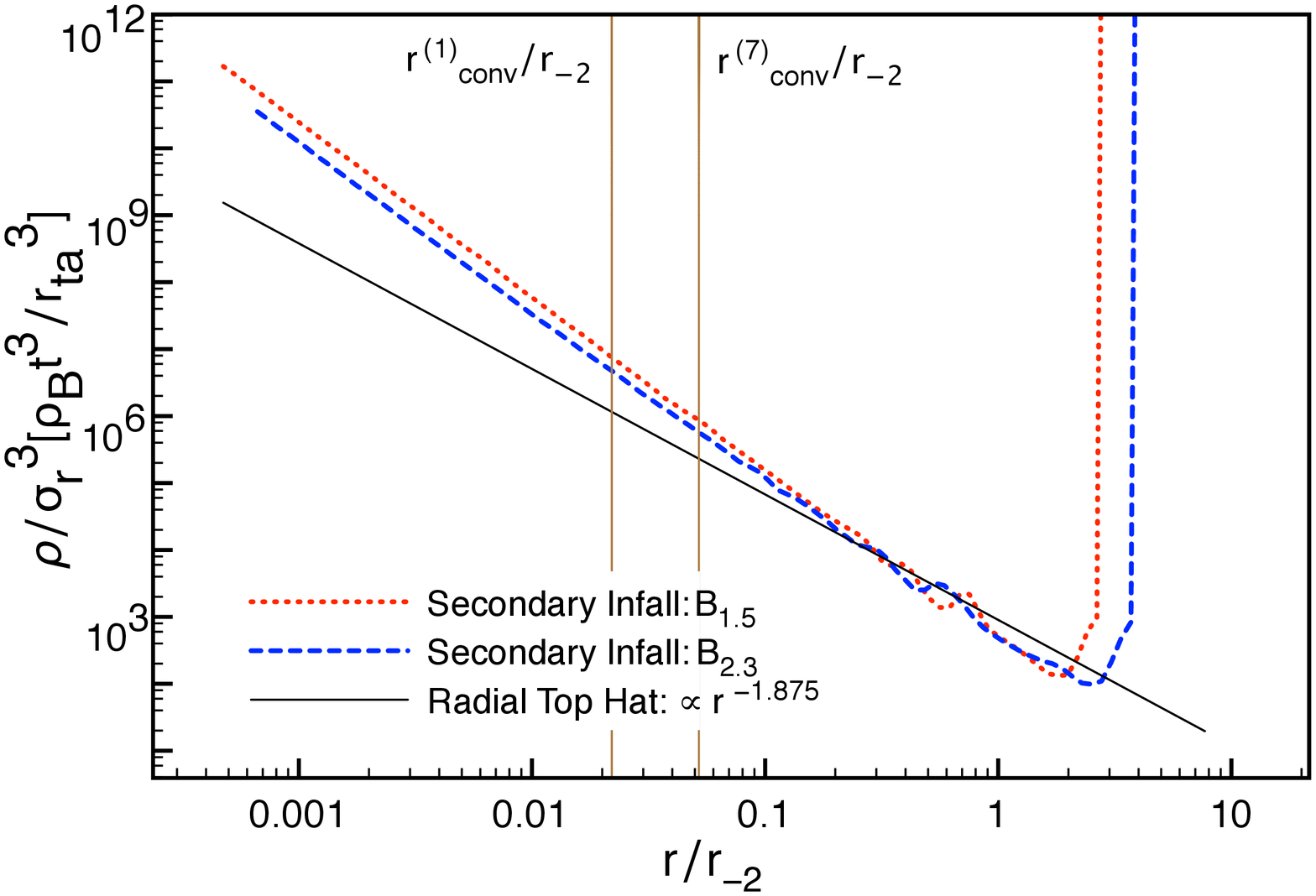}
  \end{center}
\caption{The left panels show $\rho/\sigma^3$ and $\rho/\sigma^3_r$, for a self-similar halo with model parameters  $n=.77$, $p=2n$, $B_{1.5}=0.39$, and varying $\varpi$. The numerically calculated slopes match analytic predictions. First pericenter passage $(y_0)$ for $\varpi=0$ and the virial radius $(r_v)$ are labeled for clarity. The right panels shows the smoothed pseudo-phase-space density profiles, with the radius scaled to $r_{-2}$, for model parameters that give good agreement to density profiles from galactic size simulated halos. We also plot the radial top hat prediction. The self-similar model predicts that simulations should see deviations from the radial top hat power law at $r/r_{-2}\sim 3\times10^{-2}$ for $\rho/\sigma^3$ and deviations at  $r/r_{-2}\sim 10^{-1}$ for $\rho/\sigma^3_r$. The convergence radii for the Aquarius halo Aq-A-2 \cite{Aquarius} are labeled for reference. \label{fig:PhT}}
\end{figure*}

Here we analyze the pseudo-phase-space density profiles $\rho/\sigma^3$ and $\rho/\sigma^3_r$  for galactic size halos, where $\sigma^2\equiv\sigma^2_r+\sigma^2_t$. Taylor and Navarro claimed that the pseudo-phase-space density roughly follows the power law $r^{-1.875}$ for all halos \cite{TN}.  Surprisingly, this power law matches predictions made by Bertschinger for purely radial self-similar collapse onto a spherical top hat perturbation \cite{Bert}. Taylor and Navarro's claim has been verified numerically \cite{Aquarius,Rasia,Dehnen,Faltenbacher,Vass,Wang}, however recently the highest resolution simulations have seen evidence for departures from this power law near their innermost resolved radii \cite{Ludlow}. 

Based on the analysis in Section \ref{sec:AB}, we expect the power law exponent to depend on $\{n,\varpi\}$ for $r/r_{ta}\ll y_0$. With model parameters $\{n,\varpi\}$ which give $\rho\propto r^{-1}$ for galactic size halos, the extended secondary infall model predicts $\rho/\sigma^3\propto\rho/\sigma^3_r\propto r^{-5/2}$. This is expected for a virialized halo ($\varpi>0$) with $\rho\propto r^{-1}$. For $y_0\ll r/r_{ta} \ll r/r_v$, the model predicts $\rho/\sigma^3 \propto r^{-2}$ if the radial velocity dispersion dominates and $\rho/\sigma^3 \propto r^{-1/2}$ if the tangential velocity dispersion dominates.  

In the left panels of Figure \ref{fig:PhT}, we plot $\rho/\sigma^3$ and $\rho/\sigma^3_r$ for galactic size halos with varying $\varpi$. In the right panels, we plot the smoothed pseudo-phase-space densities, with the radius scaled by $r_{-2}$, for model parameters that give good agreement with density profiles from simulated galactic size halos. In addition, we overlay the radial top hat solution. Scaling the radius to $r_{-2}$ causes the first  pericenter ($y_0$) of both models to roughly overlap, leading to less difference in the amplitude of the pseudo-phase-space density at small radii.   

The left panels show that the asymptotic slopes vary with $\varpi$. The numerically computed slopes match analytic predictions. The panels for $\rho/\sigma^3_r$ blow up at radii close to the virial radius since $\sigma_r$ vanishes. The right panels should be compared to Figure 13 of \cite{Aquarius}. The extended secondary infall model predicts that simulations of galactic size halos should see significant deviations from Taylor and Navarro's claim at $r/r_{-2}\sim3\times10^{-2}$ when analyzing $\rho/\sigma^3$ and $r/r_{-2}\sim10^{-1}$ when analyzing $\rho/\sigma^3_r$. Looking at the residuals in Figure 13 of \cite{Aquarius}, this prediction seems plausible. If higher resolution simulations do not show deviations from Taylor and Navarro's claim, then this secondary infall model would be proven incorrect since the model cannot consistently reproduce both the density and velocity profiles of simulated halos.  

As shown in Section \ref{sec:AB}, for cosmological initial conditions ($n<2$), $\rho$, $\sigma^2_t$ and $\sigma^2_r$ have power laws that are independent of initial conditions and torqueing parameters in the regime $y_0\ll r/r_{ta} \ll r_v/r_{ta}$. This implies that the pseudo-phase-space density is universal on these scales. This universality on intermediate scales may have played a role in Taylor and Navarro's initial claim. 

\begin{figure}
  \begin{center}
    \includegraphics[height = 70mm, width=\columnwidth]{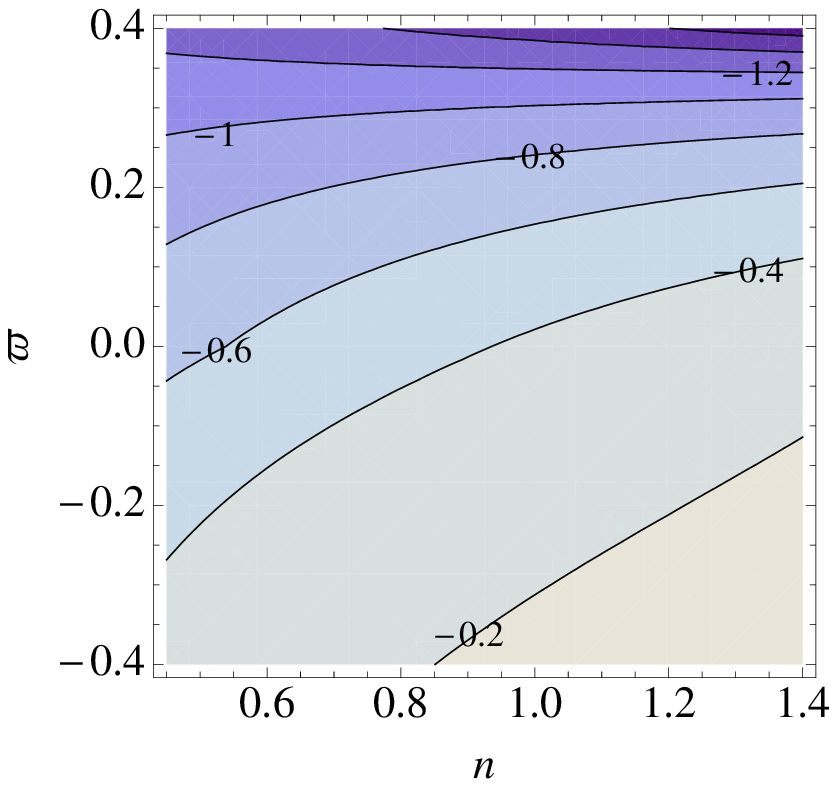}
  \end{center}
\caption{A contour plot of $d\ln (\rho/\sigma^3)/d \ln r + 1.875$, which shows the deviation in the pseudo-phase-space density power law exponent, at small scales, from the radial top hat solution. \label{fig:Contour}}
\end{figure}

Figure \ref{fig:Contour} shows the difference of the pseudo-phase-space density power law exponent from the radial top hat solution, on small scales, as a function of model parameters $n$ and $\varpi$. The range in $n$ corresponds to $10^9 < M/M_{\odot}<10^{15}$. The range in $\varpi$ ensures that all orbits are bound. According to the extended secondary infall model, positive $\varpi$ is necessary for $n>.5$ in order to have $\rho\propto r^{-1}$ on small scales \cite{zukin}. If all halos have $\rho\propto r^{-1}$ on small scales, then halos with $M>10^9M_{\odot}$ will have $\rho/\sigma^3\propto r^{-5/2}$ while halos with $M<10^9M_{\odot}$ will have pseudo-phase-space density exponents that vary with halo mass. If on the other hand, $\varpi$ is constant for all halos, then as Figure \ref{fig:Contour} shows, the power law will vary with mass.  

\section{Discussion}
\label{sec:Dis}

N-body simulations have revealed a wealth of information about the velocity profiles of dark matter halos. In an attempt to gain intuition for their results, we've used a generalized self-similar secondary infall model which takes into accounts tidal torques. The model assumes that halos self-similarly accrete radially cold mass shells. Moreover, each shell is composed of particles with the same amplitude of angular momentum. While the model is simplistic, it does not suffer from resolution limits and is much less computationally expensive than a full N-body simulation. Moreover, it is analytically tractable. Using this model we were able to analytically calculate the radial and tangential kinetic energy profiles for $r/r_{ta}\ll y_0$ and $y_0\ll r/r_{ta}\ll r/r_v$, where $y_0$ is the dimensionless radius of first pericenter passage, $r_v$ is the virial radius, and $r_{ta}$ is the current turnaround radius.

It is clear from our analysis that angular momentum plays a fundamental role in determining the velocity structure of the halo. The amplitude of angular momentum at turnaround sets the transition scale ($y_0$) between different power law behaviors in the tangential and radial kinetic energy profiles. Also, for collapsed objects today ($n<2$), $\varpi$, the parameter that quantifies how particles are torqued after turnaround, influences the slopes of both the radial and kinetic velocity dispersions at small radii. Moreover, both the amplitude of angular momentum at turnaround and $\varpi$ affect the asymptotic value of the velocity anisotropy profile at small radii.

For $\varpi<0$, the self-similar halo is not virialized on small scales since the radial and tangential kinetic energy is dominated by mass shells which have not had time to virialize. On the other hand, for $\varpi\geq0$, the halo is virialized since the dominant mass shell have had time to virialize. As shown in Paper I, $\rho\propto r^{-1}$ requires $\varpi>0$ for $M/M_{\odot}>10^9$. Hence, positive $\varpi$ is favored in order to reproduce N-body simulation density profiles. Quantifying $\varpi$ requires analyzing N-body simulations and is beyond the scope of this work. However constraining $\varpi$ with simulations will provide a test for this extended secondary infall model.

Our model predicts that the pseudo-phase-space density profile is universal on intermediate and large scales. This could potentially play a role behind the claimed universality of the pseudo-phase-space density \cite{TN}. Since we do not understand how $\varpi$ depends on halo mass, it is impossible to rule out universality on small scales, since $\varpi$ can potentially conspire to erase initial conditions. However, if galactic size halos have $\rho\propto r^{-1}$, then regardless of universality, the model predicts $\rho/\sigma^3\propto r^{-5/2}$. While hints of deviations from the radial top hat solution have been seen in recent simulations \cite{Ludlow}, higher resolution simulations are needed to better test the model.  

While our self-similar model has its clear advantages, it is also unphysical. First, all particles in a given mass shell have the same radial velocity. This leads to caustics. The same tidal torque mechanisms which cause a tangential velocity dispersion \cite{Hoyle}, should give rise to a radial velocity dispersion. Second, while qualitatively similar, the comparison of the model's predicted velocity anisotropy to N-body simulation results reveals that our treatment of angular momentum is too simplistic. The model predicts too many radial orbits at large radii and too many circular orbits at small radii. In reality, each shell is composed of a distribution of angular momentum that evolves with time. In order to properly take these two effects into account, one would need a statistical phase space description of the halo that includes sources of torque. Ma $\&$ Bertschinger provided such an analysis in the quasilinear regime \cite{MaBert}. Therefore, a natural extension of this secondary infall model, which could potentially reproduce both position and velocity space information of N-body simulations, would be to generalize Ma $\&$ Bertschinger's analysis to the nonlinear regime and impose self-similarity. 
\acknowledgments

The authors acknowledge support from NASA grant NNG06GG99G and thank Robyn Sanderson, Paul Schechter, and Mark Vogelsberger for many useful conversations.

\appendix

\section{Deriving the Consistency Relationship}

In this Appendix, we derive eq.\ (\ref{consistency}). Self-similarity imposes that the total radial (or tangential) kinetic energy at time $t$ contained within radius $r$ is given by:

\begin{equation}
K_i(r,t)=M_{ta}(t)\frac{r^2_{ta}}{t^2}{\cal{K}}_i(\lambda)
\end{equation}

\noindent where $i=\{r,t\}$ is used for shorthand to denote the radial or tangential direction. The kinetic energy also obeys the following relationship.

\begin{equation} \label{cons2}
K_i(r,t)=\frac{1}{2}\int^{M_{ta}}_0dM_*v^2_i(t,t_*)H\left[r-R(t,t_*)\right]
\end{equation}

\noindent where $dM_*, v_i(t,t_*), R(t,t_*)$ is the mass, velocity, and radius of a shell at time $t$ which turned around at $t_*$ and $H$ is the heaviside function. Since, after a short time, shells begin to oscillate on a timescale much shorter than the growth of the halo, we can replace $v_i^2$ with a time averaged version $\left\langle v^2_i\right\rangle$ and the heaviside function with a weighting that takes into account how often the shell is below $r$. More specifically, considering a shell with turnaround time $t_*$ such that $r_p(t,t_*) < r < r_a(t,t_*)$, we have:

\begin{eqnarray}
\left\langle v^2_i(t,t_*)\right\rangle &\rightarrow& \left(\int^r_{r_p}v^2_idt\right)/\left(\int^r_{r_p}dt\right) \label{vi} \\
H[r-R(t,t_*)] &\rightarrow& \left(\int^r_{r_p}dt\right)/\left(\int^{r_a}_{r_p}dt\right) \label{H}
\end{eqnarray}  

\noindent where we've left the dependence on $t_*$ implicit. Eq.\ (\ref{vi}) only averages over scales below $r$ since that is where the shell contributes to the kinetic energy. Eq.\ (\ref{H}) is identical to what is done in Fillmore and Goldreich, in order to analytically calculate the mass profile at small scales \cite{FG}.  

Using eqs.\ (\ref{vi}) and (\ref{H}), generalizing to the case where $r<r_p$ and $r>r_a$, plugging into eq. (\ref{cons2}), dividing by $K_i(r_{ta},t)$ and assuming a power law for the kinetic energy profiles in the form of eqs.\ (\ref{KinRParam}) and (\ref{KinTParam}), we reproduce the consistency equation. The equation has a proportionality constant not only because of eq.\ (\ref{IR}) but also because we do not include ${\cal{K}}_i(1)$. This overall constant does not affect the asymptotic slopes. 
  
\bibliography{VelocityPaper}

\end{document}